\documentclass[12pt]{article}
\usepackage{epsfig}
\usepackage{amssymb}

\newcommand{\be}{\begin{equation}}
\newcommand{\ee}{\end{equation}}
\newcommand{\bea}{\begin{eqnarray}}
\newcommand{\eea}{\end{eqnarray}}

\newcommand{\ve}{\varepsilon}

\newcommand{\ba}{\begin{array}}
\newcommand{\ea}{\end{array}}

\begin{document}

\begin{flushright}
ITEP-PH-4/2000\\
hep-ph/0010089
\end{flushright}

\vspace{5mm}
\centerline{\LARGE\bf The {\boldmath $V-A$} sum rules and the Operator}
\vspace{1mm}
\centerline{\LARGE\bf  Product Expansion in complex {\boldmath $q^2$}-plane }
\vspace{1mm}
\centerline{\LARGE\bf  from {\boldmath $\tau$}-decay data}
\vspace{7mm}
\centerline{\large B.L.Ioffe and K.N.Zyablyuk}
\vspace{5mm}
\centerline{e-mail: {\tt ioffe@vitep5.itep.ru, zyablyuk@heron.itep.ru}}
\vspace{3mm}
\centerline{\it Institute of Theoretical and Experimental Physics,}
\centerline{\it B.Cheremushkinskaya 25, Moscow 117259, Russia}

\begin{abstract}
The operator product expansion (OPE) for the difference of vector and axial current correlators
is analyzed for complex values of momentum $q^2$. The vector and axial spectral functions,
taken from hadronic $\tau$-decay data, are treated with the help of Borel, Gaussian and
spectral moments sum rules. The range of applicability, advantages and disadvantages 
of each type are discussed. The general features of OPE are confirmed by the data.
The vacuum expectation values of dimension 6 and 8 operators
are found to be $O_6=-(6.8\pm 2.1)\times 10^{-3} \, {\rm GeV}^6$, 
$O_8=(7\pm 4)\times 10^{-3} \, {\rm GeV}^8$.
\end{abstract}

\section{Introduction}

Precise measurements of vector $V$ and axial $A$ spectral functions in $\tau$-decay have been
recently performed by ALEPH \cite{ALEPH2} and OPAL \cite{OPAL} collaborations.
Define the polarization operators of hadronic currents:
\be
\label{pol0}
\Pi_{\mu\nu}^U(q)\,=\,
i\!\int e^{iqx}<TU_\mu(x) U_{\nu}(0)^\dagger> dx\,=\,\left(q_\mu q_\nu -g_{\mu\nu} q^2\right)
\Pi_U^{(1)}(q^2)+ q_\mu q_\nu \Pi_U^{(0)}(q^2)  \; ,
\ee
$$
{\rm where} \qquad U=V,A \, ;   \qquad 
V_\mu=\bar{u}\gamma_\mu d  \; , \qquad 
A_\mu=\bar{u}\gamma_\mu\gamma_5 d \; .
$$
The imaginary parts of the correlators are the so-called spectral functions ($s=q^2$),
\be
v_1/a_1(s)=2\pi\,{\rm Im}\,\Pi^{(1)}_{V/A}(s+i0) \; , \qquad
a_0(s)=2\pi\,{\rm Im}\,\Pi^{(0)}_{A}(s+i0) \; . 
\ee
which have been measured from hadronic $\tau$-decays
for $0<s<m_\tau^2$. The spin-0 axial spectral function  $a_0(s)$ is basically saturated by 
$\tau\to \pi\nu_\tau$ channel, which gives $\delta$-function. It will not be considered here. 

In this paper the experimental data for $v_1-a_1$ will be used to determine numerical 
values of the quark condensates in QCD. An early attempt to realize such programm was performed by
Eidelman, Vainstein and Kurdadze \cite{EVK} using $e^+e^-$ annihilation data, but the experimental
errors were rather large and the result not very conclusive. Also, higher order condensates and
higher order perturbative corrections were not included in the analysis. More recent analysis
\cite{G} of $e^+e^-$ annihilation data demonstrates, that the spread of the values of the quark and
gluon condensates is larger than found in \cite{EVK}. Therefore the consideration of the problem
based on new precise $\tau$-decay data is reasonable.

The spin-1 parts $\Pi^{(1)}_V(q^2)$ and $\Pi^{(1)}_A(q^2)$ are
analytical functions in the complex $q^2$-plane with a cut along the right semiaxes starting from
the threshold of the lowest hadronic state: $4m_\pi^2$ for $\Pi^{(1)}_V$ and 
$9m_\pi^2$ for $\Pi^{(1)}_A$. The latter has a kinematical pole at $q^2=0$. This is a 
specific feature of QCD, which follows from the chiral symmetry in the limit of massless 
$u,d$-quarks and its spontaneous violation. Indeed, in this limit the axial current is conserved and
there exists a massless Goldstone boson, namely the pion. Its contribution to the axial polarization 
operator is given by:
\be
\label{pola1}
\Pi_{\mu\nu}^A(q)_\pi\,=\,f_\pi^2\left( \,g_{\mu\nu}\,-\,{q_\mu q_\nu \over q^2}\, \right) \; ,
\ee
where $f_\pi$ is the pion decay constant, $f_\pi=130.7 \; {\rm MeV}$ \cite{PDG}. When the quark
masses are taken into account, then in the first order of quark masses or, what is equivalent, in
$m_\pi^2$, eq.~(\ref{pola1}) gets modified:
\be
\label{pola2}
\Pi_{\mu\nu}^A(q)_\pi\,=\,f_\pi^2\left( \,g_{\mu\nu}\,-\,{q_\mu q_\nu \over q^2-m_\pi^2}\, \right) \; .
\ee
It can be decomposed in the tensor structures of (\ref{pol0}):
\be
\label{pola3}
\Pi_{\mu\nu}^A(q)_\pi\,=\,-\,{f_\pi^2\over q^2}\left( \, q_\mu q_\nu - g_{\mu\nu}q^2\, \right)
-\,{m_\pi^2\over q^2}\, q_\mu q_\nu \, {f_\pi^2\over q^2- m_\pi^2} 
\ee
According to this equation the residue at the kinematical pole is equal to $-f_\pi^2$. The accuracy
of this statement is of order of the chiral symmetry violation in QCD, $\sim m_\pi^2/m_\rho^2$, 
where $m_\rho$ is characteristic hadronic scale (say, $\rho$-meson mass) \cite{GL}
(e.g. a subtruction term $\sim g_{\mu\nu} f_\pi^2 m_\pi^2/m_\rho^2$ can be added to 
$\Pi^A_{\mu\nu}$).

\begin{figure}[tb]
\hspace{35mm}
\epsfig{file=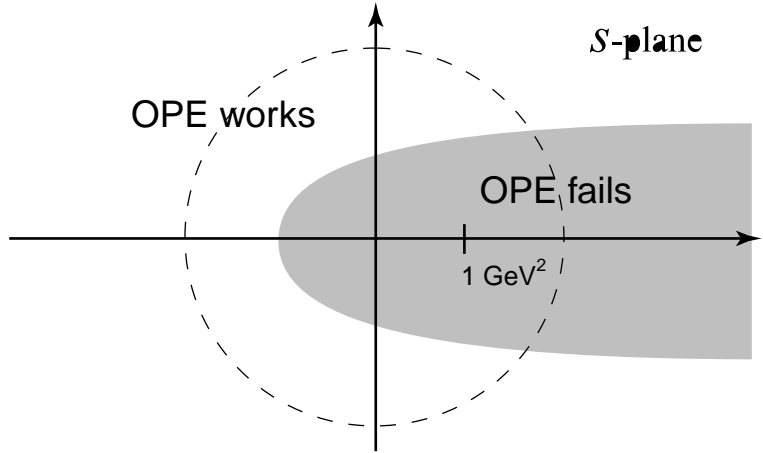}%, height=50mm, width=70mm}
\caption{Domain of OPE validity}
\label{opereg}
\end{figure}

The difference $\Pi_V-\Pi_A$ is of particular interest, since in QCD it does not have any perturbative
contribution in the limit of massless quarks. 
We use the analytical properties of $\Pi^{(1)}_V(s)$ and $\Pi^{(1)}_A(s)$ in the complex $s$-plane in
order to construct the sum rules for $\Pi_V-\Pi_A$ valid at large $|s|$. At large $|s|$ the operator
product expansion (OPE) takes place
\be
\label{ope1}
\Pi_V^{(1)}(s)-\Pi_A^{(1)}(s)\,=\,\sum_{D\ge 4} \,{O^{V-A}_D \over
(-s)^{D/2} } \left( \,1\,+\,c_D {\alpha_s\over \pi}\, \right)\,=\, 
\sum_{D\ge 4} \,{O_D \over (-s)^{D/2} } \; ,
\ee
where $O^{V-A}_D$ are the vacuum averages of local operators, constructed from quark
and gluon field. In what follows the operators $O_D$ without index $V-A$ include the
radiative corrections $O_D=O_D^{V-A}(1+c_D\alpha_s/\pi)$.
Higher order perturbative corrections to $O_D^{V-A}$, as well as the terms $\sim m_{u,d}^2$ are
neglected. One may expect, that OPE is valid in the whole complex $s$-plane, except for the
domain of small $|s|$ and near positive real semiaxes (see Fig.~\ref{opereg}).

\begin{figure}[tb]
\hspace{5mm}
\epsfig{file=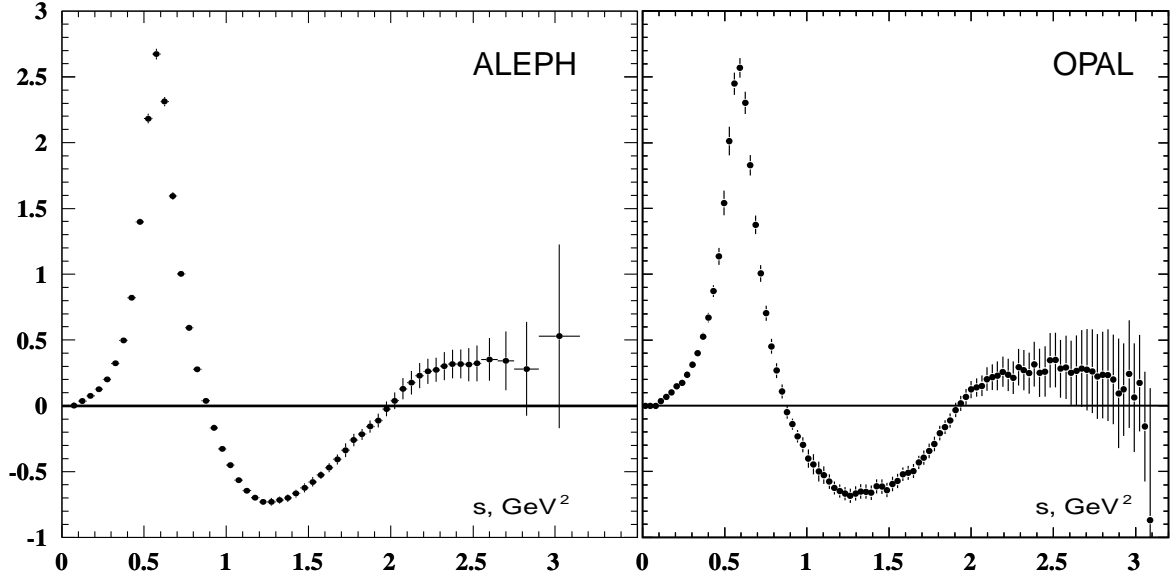}%, height=75mm, width=150mm}
\caption{The measured difference $v_1(s)-a_1(s)$. Figures from \cite{ALEPH2} and \cite{OPAL}.}
\label{vma_exp}
\end{figure}

The measured difference of the spectral functions $v_1(s)-a_1(s)$ is shown in Fig.~\ref{vma_exp}. 
In this paper we use the ALEPH data, since the files with invariant mass spectra and
correspondent covariance matrices are publicly available.

The operators $O^{V,A}_D$ have been computed up to dimension $D=6$ in \cite{SVZ}. 
The earlier calculations of $O^V_8$ have been done in \cite{BG}--\cite{DS}, but there
are some discrepancies in the results. We have recalculated $O_8^{V,A}$ (see Appendix).
In the calculation of $O_6^{V-A}$ and $O_8^{V-A}$ the factorization hypothesis, i.e. the
saturation by intermediate vacuum state, is assumed. As shown in Appendix, there is
an ambiguity in the  factorization of $D=8$ operators among the terms 
of order $N_c^{-2}$; they are neglected here. The results are:
\bea
O^{V-A}_4 & = & 2 \,(m_u +m_d)\,<\bar{q}q>  \; = \; -\, f_\pi^2 m_\pi^2
\label{o4}  \\
O^{V-A}_6 & = & 2\pi \alpha_s \left<\,
(\bar{u}\gamma_\mu\lambda^a d)(\bar{d}\gamma_\mu \lambda^a u) -
(\bar{u}\gamma_5\gamma_\mu\lambda^a d)(\bar{d}\gamma_5\gamma_\mu \lambda^a u)\, \right>
 \nonumber \\
 & = & -\,{64\pi\alpha_s\over 9} <\bar{q}q>^2  \label{o6vma} \\
O^{V-A}_8 & = &  8\pi \alpha_s \, m_0^2 <\bar{q}q>^2 \; .
\label{o8vma}
\eea
The definition of $m_0^2$ is given in Appendix, we assume the isotopic symmetry
among the quark condensates: $<\bar{u}u>=<\bar{d}d>=<\bar{q}q>$.

Let us discuss what is known about the vacuum averages $O_D^{V-A}$. 
The numerical value $O_4^{V-A}=-f_\pi^2 m_\pi^2=-3.4\times 10^{-4} \;
{\rm GeV}^4$ is very small and in almost all cases can be ignored. The quark condensate
$<\bar{q}q>$ can be found from Gell-Mann-Oakes-Renner low energy theorem \cite{GOR}:
\be
\label{qc1}
<\bar{q}q>\,=\,-\,{1\over 2}\,{f_\pi^2 m_\pi^2 \over m_u+m_d}
\ee
At standard values (see e.g. \cite{IKL}) $m_u=4.2 \; {\rm MeV}$, $m_d=7.5 \; {\rm MeV}$ we have
\be
\label{qc2}
<\bar{q}q>\,=\,-\,1.4 \times 10^{-2} \; {\rm GeV}^3
\ee
The value of $<\bar{q}q>$ depends on the normalization point $\mu^2$ and it is unclear to which
normalization point it refers. In recent analysis of QCD sum rules for proton \cite{I} the  same
numerical value as (\ref{qc2}) was found at the point $\mu^2=1\; {\rm GeV}^2$. Using this value
and $\alpha_s(1\; {\rm GeV}^2)=0.5$, which follows from $\alpha(m_Z^2)=0.119$ by using 
three loop QCD renormalization group evolution, we get for renorminvariant quantity:
\be
\label{qc3}
\alpha_s \, <\bar{q}q>^2\,=\,1.0 \times 10^{-4} \; {\rm GeV}^6
\ee
Here, however, we have to be careful. In QCD sum rule analysis \cite{I} no $\alpha_s$ corrections were
accounted. They may result in $20-30\%$ uncertainty in $<\bar{q}q>_{1\, {\rm GeV}^2}$.
Taking (\ref{qc3}), we get for $O_6^{V-A}$:
\be
\label{op6e1}
O_6^{V-A}\,=\,-\,2.2\times 10^{-3} \; {\rm GeV}^6
\ee
The value of $m_0^2$ was found in \cite{BI} from the analysis of QCD sum rules for baryons:
\be
\label{m01}
m_0^2\,=\,0.8\pm 0.2 \; {\rm GeV}^2
\ee
The substitution of (\ref{qc3}) and (\ref{m01}) in (\ref{o8vma}) gives:
\be
\label{op8e1}
O_8^{V-A}\,=\,2\times 10^{-3} \; {\rm GeV}^8
\ee

Perturbative $\alpha_s$ corrections were calculated for the contribution of 
 $D=4$ \cite{CGS} and $D=6$ \cite{LSC, AC} operators. The correction to $O_4^{V-A}$ is
 $c_4=4/3$; it increases the effective value of the operator $O_4$ on $20\%$.
Concerning $O_6$,  two essentially different values have been obtained:
$c_6=247/48$ in \cite{LSC} and $c_6=89/48$ in \cite{AC}
\footnote{There is also the logarithmical correction $ (\alpha_s/4\pi)\ln{(s/\mu^2)}$ to the operator $O_6$,
which was not included in (\ref{ope1}). However this term is small for physically reasonable 
values of the scale $\mu^2$.}. In \cite{AC} it was argued, that the
last one is more reliable, since in its calculation the correct treatment of $\gamma_5$ in
dimensional regularization scheme was done and more plausible vacuum saturation of 4-quark
operators was performed. If we take $c_6=89/48$, put $\alpha_s(1\, {\rm GeV}^2)=0.5$ and neglect the
$q^2$ dependence, then we get for the effective operator:
\be
\label{op6e2}
O_6\,=\,-\,3\times 10^{-3} \; {\rm GeV}^6
\ee
In leading order $O_8$ weakly depends on the normalization point. So we will consider 
$O_8=O_8^{V-A}$ as effective $D=8$ operator with $\alpha_s$ correction included.

Our goal is to find $O_6$ and $O_8$ from $\tau$ decay data and compare them with (\ref{op6e2})
and (\ref{op8e1}). Higher order operators with $D\ge 10$ also contribute to OPE (\ref{ope1}).
OPE is an asymptotic series. The comparision of numerical values (\ref{op6e1}) and
(\ref{op8e1}) indicates, that at $|s|=1\,{\rm GeV}^2$ this series starts to diverge at $D=8$
(the same conclusion $|O_6|\sim |O_8|$ in GeV follows also from our final result). Therefore in order to
get reliable results we have to go to higher $|s|$ or to improve the convergence of the series.
In order to estimate the error in the $O_{6,8}$ determination we will accept the conservative 
assumption, that $O_D$ measured in $(\rm GeV)^D$ increase starting from $D=10$,
for instance $|O_{10}|\sim 2|O_6|$.

\section{Moments sum rules}

The dispersion relation or Borel transformation requires the knowledge of the spectral 
functions for all $s$. Although the vector function $v_1(s)$ within isotopic symmetry 
can be found for $s>m_\tau^2$ from $e^+e^-$ annihilation, the precision is low, since
the experimental analysis involves the states with 6 mesons and more.
The axial-vector function $a_1(s)$ is not known beyond this point at all.

The technique of the spectral moments \cite{DP} used for the evaluation 
of hadronic $\tau$-decay branching ratio does not need this information. 
The following moments are computed:
\bea
M^{kl}(s_0) & = & {1\over 2\pi^2}\int_0^{s_0} {ds\over s_0} \left(1-{s\over s_0}\right)^k
\left({s\over s_0}\right)^l (v_1-a_1)(s)   \label{mome} \\ 
 & = & {i\over 2\pi} \oint_{|s|=s_0} {ds\over s_0} \left(1-{s\over s_0}\right)^k
\left({s\over s_0}\right)^l (\Pi_V^{(1)} - \Pi_A^{(1)})(s)   
\,=\,(-)^l\sum_{j=0}^k C^k_j \,{O_{2(l+j+1)}\over s_0{}^{l+j+1}} \; ,
\label{momt}
\eea
$C^k_j$ are binomial factors (we take $O_2=f_\pi^2$ here). 
For $s_0<m_\tau^2 $ the moments can be computed from experimental data.
In the equation (\ref{momt}) the integral goes counterclockwise over  the circle with radius $s_0$.

In principle one can find all operators $O_D$ in this way. Nevertheless, for $k<2$ the
experimental error is very high, so the number of independent moments in (\ref{momt})
which can be computed with desirable accuracy is less, than
the number of unknown operators. Consequently
we have to neglect the contribution of higher dimensional operators, introducing thereby 
a theoretical uncertainty. 

In order to find the operators up to $D=8$, one should compute four independent moments
$M^{kl}$. The experimental error is large for small $k$ and large $l$. The theoretical
uncertainty grows with $k+l$, since unknown operators from $O_{10}$ to $O_{2(k+l+1)}$
are involved. Although the experimental error could be in acceptable range, the
result depends on particular set of moments. 

\begin{figure}[tb]
\hspace{5mm}
\epsfig{file=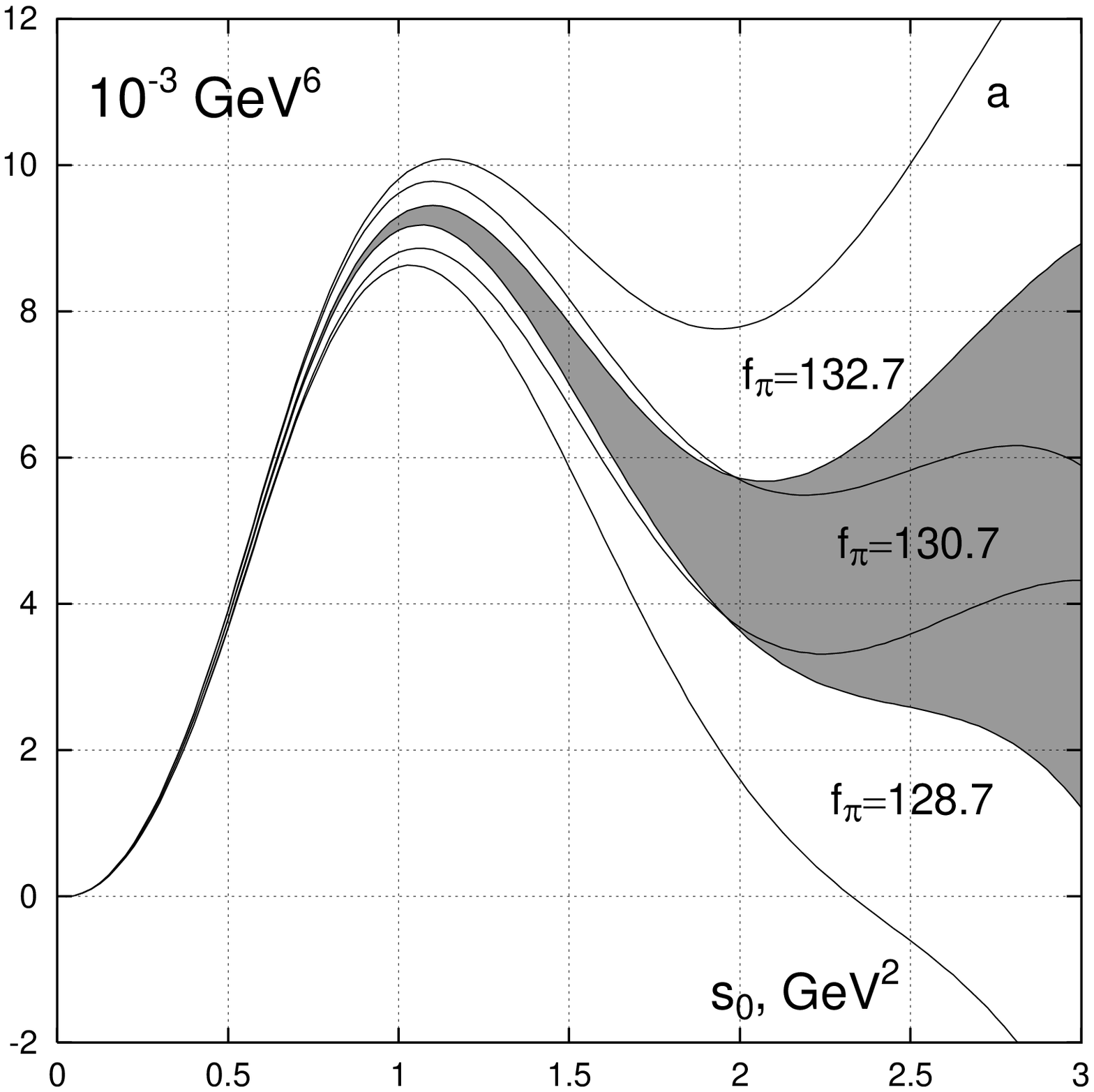, height=65mm, width=70mm} \hspace{10mm}
\epsfig{file=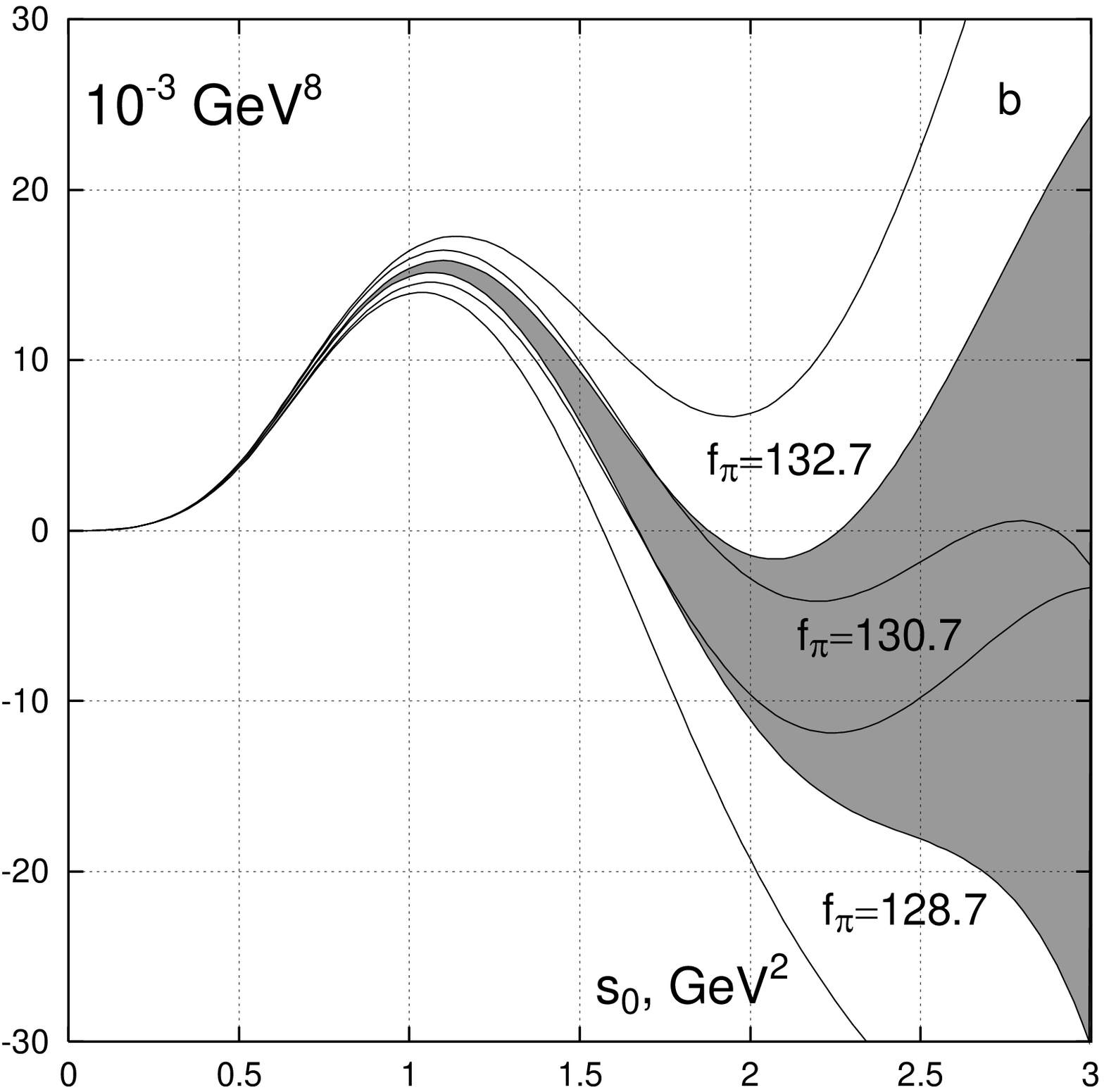, height=65mm, width=70mm}
\caption{Right hand sides of the equation (\ref{o6m}) for $n=3$ (a) and $n=4$ (b).}
\label{o6o8fig}
\end{figure}

On the other hand, $f_\pi^2$ and $O_4$ are known from other data with high accuracy.
One may use this information and moments with $k=2$ in order to find the
operator of dimension 6 and higher:
\be
(-)^n\,O_{2n} \,  = \, s_0^n \left[\, -\,\sum_{l=0}^{n-3}(n-l-2)M^{2l}(s_0)\, 
+ \,(n-2)\,{f_\pi^2\over s_0} \,+\,(n-1)\,{O_4\over s_0^2}\, 
 \right]  \; , \qquad n\ge 3 \, .  \label{o6m}
\ee
Provided that the OPE (\ref{ope1}) works, the r.h.s. of this equation
should not depend on $s_0$. It is plotted versus $s_0$ in the Fig.~\ref{o6o8fig}a,b for
$n=3$ and $n=4$ respectively. According to these figures the operator $O_6$ can be estimated as 
$-(5\pm 3)\times 10^{-3} \, {\rm GeV}^6$, while the operator $O_8$ is even remotely 
does not look as a constant.  The uncertainty in the determination of $f_\pi$ strongly
affects the result. In the Figs \ref{o6o8fig}a,b we have plotted the operators for 3 different cases:
the central value $f_\pi =130.7 \, {\rm MeV}$ and with $\pm 1.5\%$ excess.

The reason of this failure is the invalidity of the expansion (\ref{ope1}) in all complex $s$-plane.
In the moments (\ref{momt}) the integral over the circle crosses the area
where OPE does not work (see Fig.~\ref{opereg})
and it is questionable whether this contribution is suppressed enough by the factor $(1-s/s_0)^k$. 
As Fig.~\ref{o6o8fig} demonstrates, this is true only for the radius of the circle greater than 
$\sim 2 \, {\rm GeV}^2$.

In principle eq.~(\ref{o6m}) can be used for $n\ge 5$, but the experimental error in this
case is so high that it does not allow us to extract any reliable information about the values
of $D\ge 10$ operators.

\section{Borel sum rules}

The Borel sum rules can be considered at complex values of $s$.
Represent $\Pi^{(1)}_V-\Pi^{(1)}_A$ via unsubtructed dispersion relation
\be
\Pi^{(1)}_V(s)-\Pi^{(1)}_A(s)\,=\,{1\over 2\pi^2}\int_0^\infty {v_1(t)-a_1(t)\over t-s} \, dt\,+
\,{f_\pi^2\over s}
\label{dr}
\ee
The last term in the r.h.s. is the contribution of the kinematic pole.
Let us substitute the OPE (\ref{ope1}) in the l.h.s. of (\ref{dr}). Consider $s$ in the complex plane
$s=s_0e^{i\phi}$ ($\phi=0$ on the upper side of the cut) and perform Borel (Laplace) transformation
of (\ref{dr}) by $s_0$. The real and imaginary parts give us the following sum rules:
\be
\label{bre}
\int_0^\infty 
\exp{\!\left({s\over M^2}\cos{\phi}\right)}\cos{\!\left({s\over M^2}\sin{\phi}\right)}
(v_1-a_1)(s)\,{ds\over 2\pi^2} \, = \, f_\pi^2+\,\sum_{k=1}^\infty 
(-)^k {\cos{(k\phi)} \,O_{2k+2}\over k!\, M^{2k}} 
\ee
\be
\label{bim}
\int_0^\infty 
\exp{\!\left({s\over M^2}\cos{\phi}\right)}\,\sin{\!\left({s\over M^2}\sin{\phi}\right)}
(v_1-a_1)(s)\,{ds\over 2\pi^2 M^2} \, =  \,\sum_{k=1}^\infty 
(-)^k {\sin{(k\phi)} \,O_{2k+2}\over k!\, M^{2k+2}} 
\ee
The expression in the exponent is negative for $\pi/2<\phi<3\pi/2$. Since 
eq.~(\ref{bre}) is symmetric and eq.~(\ref{bim}) is antisymmetric
in the lower half plane, it is enough to analyze the region $\pi/2<\phi<\pi$. 

\begin{figure}[tb]
\hspace{5mm}
\epsfig{file=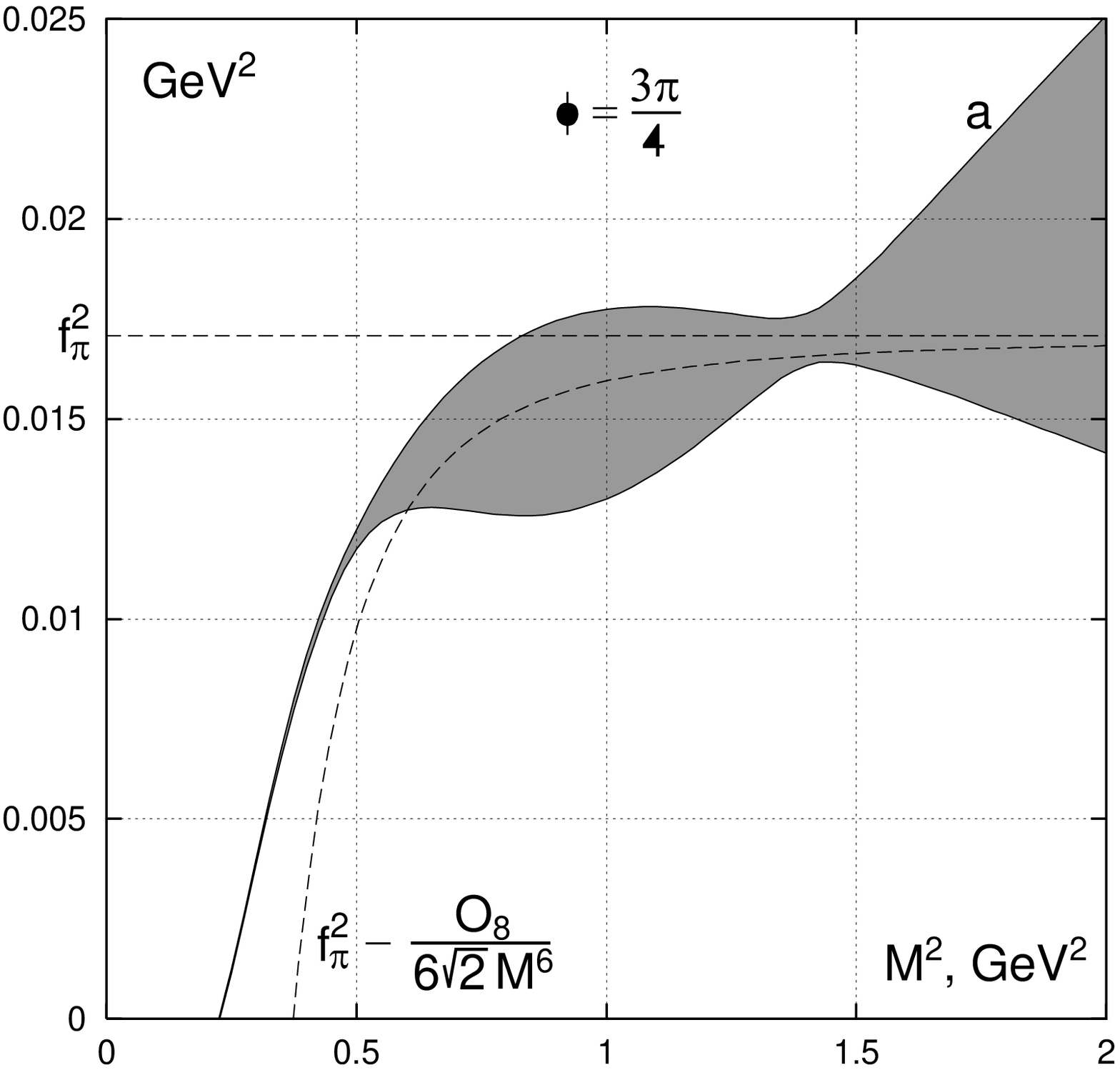, height=65mm, width=70mm} \hspace{10mm}
\epsfig{file=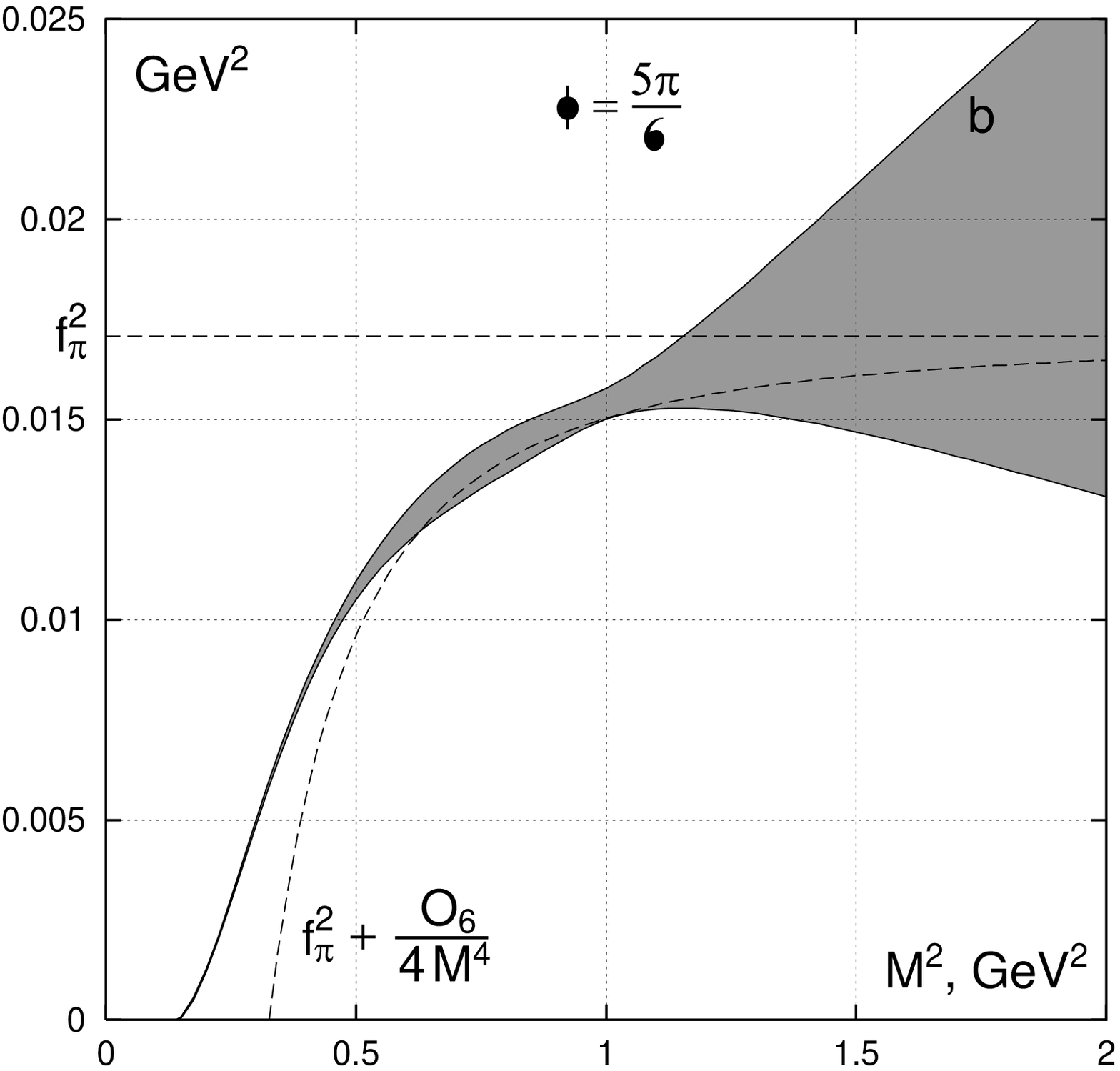, height=65mm, width=70mm}
\caption{Left hand side of (\ref{bre}). Dash lines display OPE prediction with operators
equal to the central values of (\ref{borbest}).}
\label{borc}
\end{figure}

\begin{figure}[tb]
\hspace{5mm}
\epsfig{file=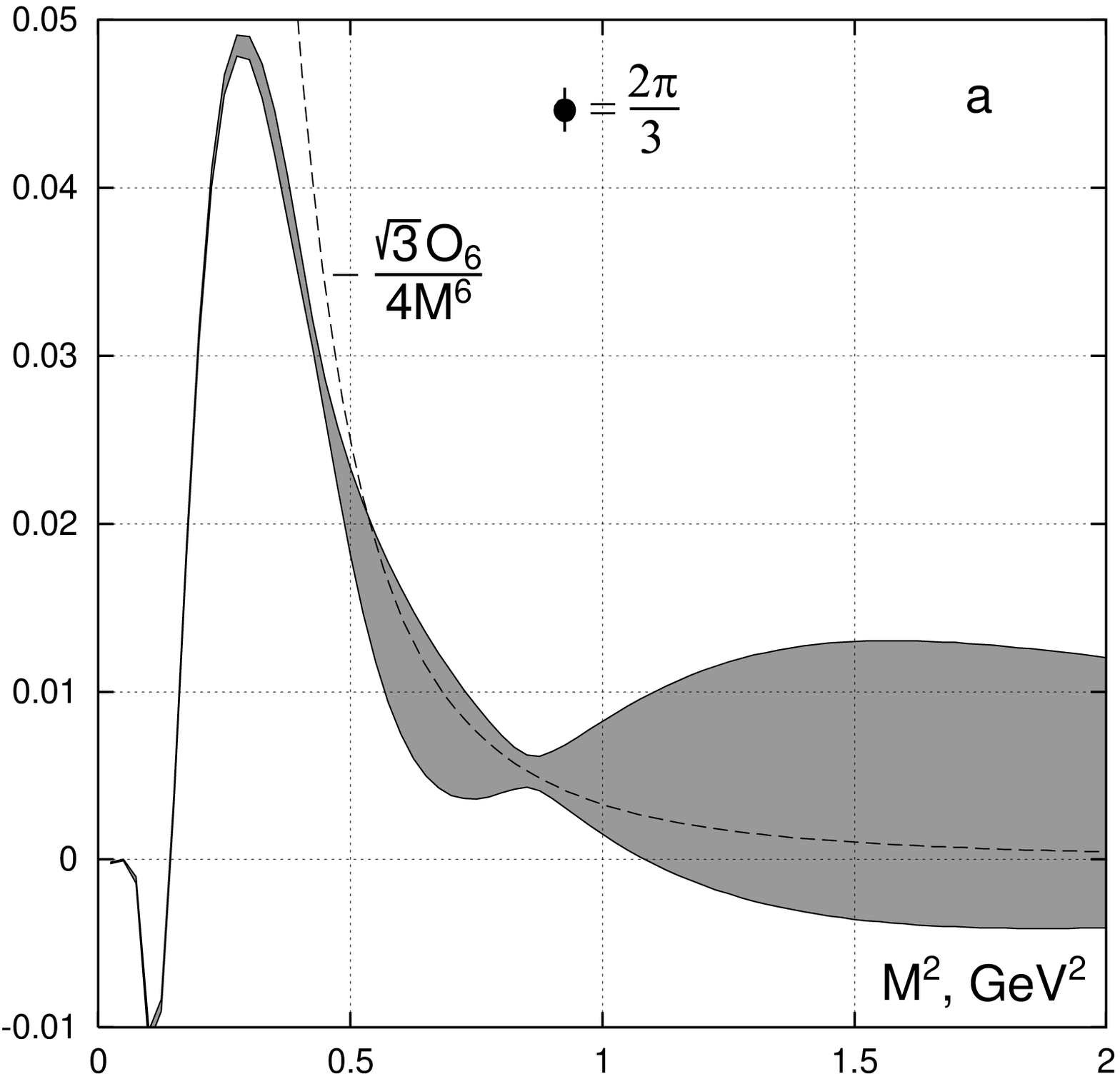, height=65mm, width=70mm} \hspace{10mm}
\epsfig{file=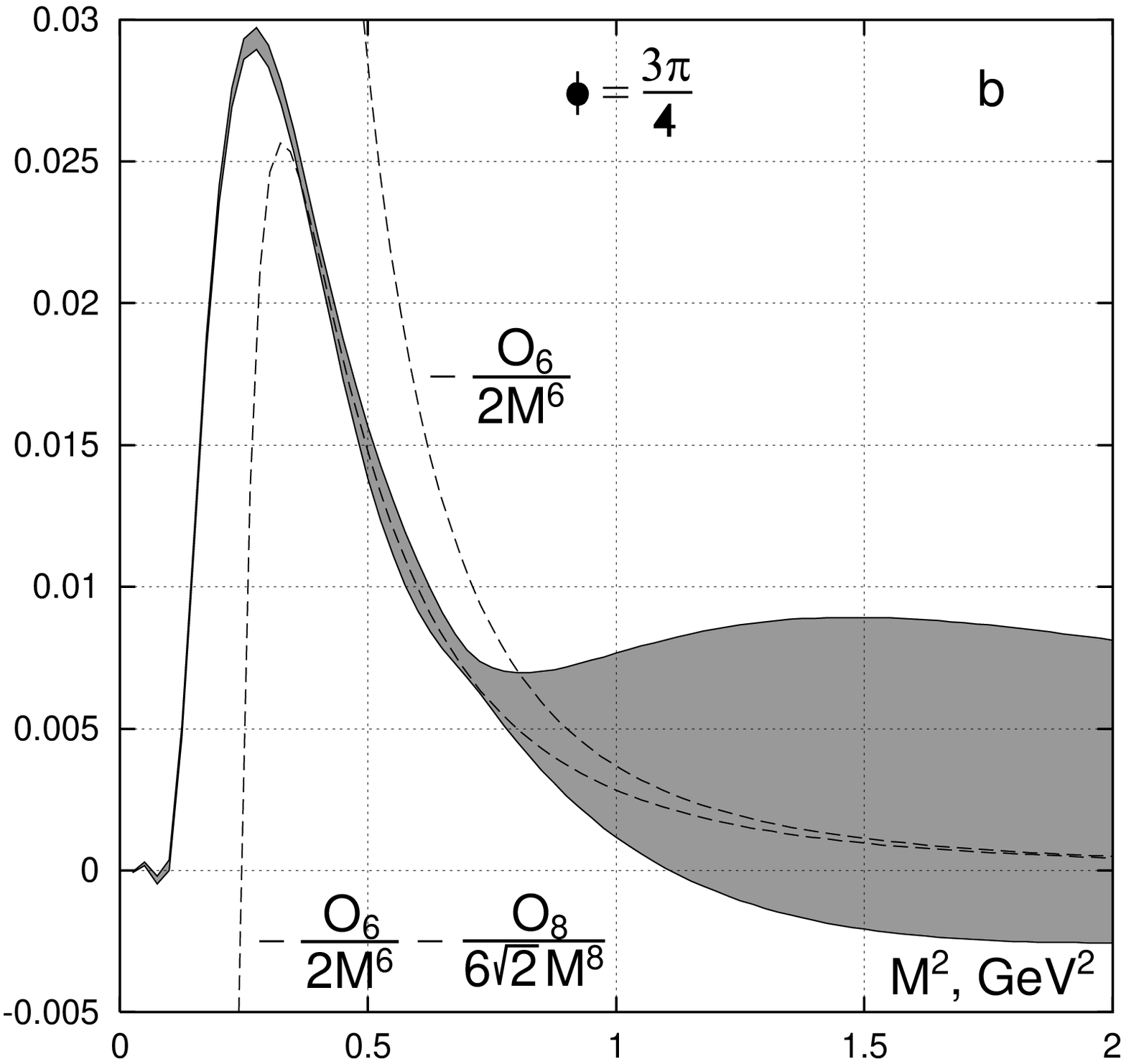, height=65mm, width=70mm}
\caption{Left hand side of (\ref{bim}). Dash lines display OPE prediction with operators
equal to the central values of (\ref{borbest}).}
\label{bors}
\end{figure}

At certain angles the contribution of some operators vanishes. This fact
can be used to separate the operators from each other. In particular, 
eq.~(\ref{bre}) with $\phi=3\pi/4$ and eq.~(\ref{bim}) with $\phi=2\pi/3$
do not contain the operator of dimension 8.
For $\phi=5\pi/6$ the operator $O_6$ disappears from the eq.~(\ref{bre}) and mainly the 
operator $O_8$ contributes to the excess over $f_\pi^2$. All these cases are shown in 
Figs \ref{borc},\ref{bors}.
We also show eq.~(\ref{bim}) for $\phi=3\pi/4$, where the contributions of the operators
$O_6$ and $O_8$ are comparable. Thin areas on the graphs are just because the sin or cos
for particular $\phi$ and $M^2$  has zero at $s=m_\tau^2$ , where the experimental error is high. 


Borel sum rules have serious advantage, since the operators of higher dimensions are factorially
suppressed. This allows one in the sum rules to go from above up to $M^2\approx 1\, {\rm GeV}^2$
and even lower in some cases. But they have also a disadvantage: at $M^2> 1\, {\rm GeV}^2$
the upper tail of the integrals in the l.h.s.'s of (\ref{bre}), (\ref{bim}) are not suppressed enough.
But, luckely, the oscillating factors in the l.h.s.'s of (\ref{bre}), (\ref{bim}) help in some cases as
can be seen from Figs \ref{borc},\ref{bors}. We exploit this fact.
 
Let us look first at eq.~(\ref{bre}) at $\phi=5\pi/6$. The r.h.s. of (\ref{bre}) is equal to:
\be
f_\pi^2\,+\,{\sqrt{3}\over 2}\,{O_4 \over M^2}\,+\,{1\over 4}\,{O_6\over M^4}\,-
\,{1\over 48}\,{O_{10}\over M^8} \; ,
\label{bre5pi6}
\ee 
higher orders are discarded. As seen from Fig \ref{borc}b at $M^2=0.8\, {\rm GeV}^2$, the
deviation from $f_\pi^2$ is definitely outside the limit of errors. The second term in (\ref{bre5pi6})
is small $\approx -3.0 \times 10^{-4}\; {\rm GeV}^2$. The main contribution comes from the operator
$O_6$, since $O_{10}$ contribution is strongly suppressed. If we neglect it, then one gets from
Fig \ref{borc}b:
\be
O_6\,=\,-(6.3\pm 1.4)\times 10^{-3} \; {\rm GeV}^6
\label{o6nb2}
\ee
Possible contribution of $O_{10}$ (at $O_{10}\sim 2|O_6|$) and $3\%$ uncertainty in $f_\pi^2$
increases the error to $2.5\times 10^{-3} \, {\rm GeV}^6$ (the errors are added in quadratures) and 
finally we get from eq.~(\ref{bre}) at $\phi=5\pi/6$:
\be
O_6\,=\,-(6.3\pm 2.5)\times 10^{-3} \; {\rm GeV}^6
\label{o6nb3}
\ee
This value can be checked by considering the sum rule (\ref{bim}) at $\phi=2\pi/3$ (Fig \ref{bors}a).
The r.h.s. of (\ref{bim}) reads:
\be
-\,{\sqrt{3}\over 2}\,{O_4 \over M^4} \,-\, {\sqrt{3}\over 4}\,{O_6\over M^6}\,+
\,{\sqrt{3}\over 48}\,{O_{10}\over M^{10}}
\label{bim2pi3}
\ee
The most suitable $M^2$ is in the region of the isthmus in the experimental errors area,
$M^2\approx 0.85 \, {\rm GeV}^2$. Here, according to Fig \ref{bors}a, the l.h.s. of (\ref{bim})
is $(5.3\pm 1.0)\times 10^{-3}$ and we have from (\ref{bim2pi3})
\be
O_6\,=\,-(6.8\pm 2.1)\times 10^{-3} \; {\rm GeV}^6
\label{o6nb4}
\ee
in agreement with (\ref{o6nb2}) (the error from $O_{10}$ is included). 

Let us try to find the value of the operator $O_8$. In (\ref{bre}) at $\phi=3\pi/4$ the 
contribution of $O_6$ vanishes and in the r.h.s. we get:
\be
f_\pi^2\,+\,{1\over \sqrt{2}}\,{O_4\over M^2}\,-\,{1\over 6\sqrt{2}}\,{O_8\over M^6}
\,-\,{1\over 24}\,{O_{10}\over M^8}
\ee
The most appropriate domain of $M^2$ is the area of small $M^2$, where the deviation 
from $f_\pi^2$ is remarkable. At the assumption $|O_{10}|\sim 2|O_6|$ the minimal squared error
is achieved at $M^2=0.6\, {\rm GeV}^2$
\be
O_8\,=\,(6\pm 8)\times 10^{-3}\; {\rm GeV}^8 \; ,
\ee
which gives us only the upper limit of $O_8<14\times 10^{-3}\, {\rm GeV}^8$. Similar upper limit
follows from consideration of large $M^2\approx 1.4 \, {\rm GeV}^2$, where the contribution
of $O_{10}$ operator is small and experimental error dominates.

Consider finally the eq.~(\ref{bim}) at $\phi=3\pi/4$, where both operators $O_6$ and $O_8$
contribute (Fig \ref{bors}b). This value of $\phi$ has the advantage, that $O_{10}$ operator
disappear from the r.h.s. of (\ref{bim}), which becomes:
\be
-\,{1\over \sqrt{2}}\,{O_4\over M^4}\,-\,{1\over 2}\,{O_6\over M^6}
\,-\,{1\over 6\sqrt{2}}\,{O_8\over M^8}\,+\,{1\over 120\sqrt{2}}\,{O_{12}\over M^{12}}
\label{bim3pi4}
\ee
The small numerical factor in front of $O_{12}$ operator allows one to go to low values of $M^2$,
where the experimental errors are small. We choose $M^2=0.65\, {\rm GeV}^2$. Then, even if
$O_{12}\sim 5 |O_6|$, its contribution to (\ref{bim3pi4}) is small. At $M^2=0.65\, {\rm GeV}^2$
the data give the value $(8.5\pm 0.6)\times 10^{-3}$ for the expression (\ref{bim3pi4}).
The substitution of $O_6=-(6.8\pm 2.1)\times 10^{-3}\,{\rm GeV}^6$ given by (\ref{o6nb4}) results to:
\be
O_8\,=\,(7\pm 7)\times 10^{-3} \; {\rm GeV}^8
\ee
(The possible error from $O_{12}$ contribution is accounted at $|O_{12}|\sim 5 |O_6|$,
all errors are added quadratically.) Again, only the upper limit.

More definite result for $O_8$ can be obtained if we accept more optimistic assumption,
that the magnitudes of $O_{10,12}$ operators in GeV are of the same order as $|O_6|$. Then
the error of $O_6$ in eq.~(\ref{o6nb4}) 
is reduced to $1.6 \times 10^{-3}$ and in eq.~(\ref{bim}) at $\phi=3\pi/4$
one may go down to $M^2=0.4 \; {\rm GeV}^2$ to minimize  the total error. In this case
our best values from Borel sum rules are:
\be
O_6\,=\,-(6.8\pm 1.6)\times 10^{-3} \; {\rm GeV}^6 \; , \qquad
O_8\,=\,(7.2\pm 3.4)\times 10^{-3} \; {\rm GeV}^8
\label{borbest}
\ee
These results must be taken with a certain care, sine the errors may be underestimated:
at such low $M^2$ there could be some terms, not given by OPE (e.g. of exponential
type, $\exp{(-\rho\sqrt{-s})}$).

\section{Gaussian sum rules}

In Borel sum rules the spectral functions are integrated with the weight function $e^{-s/M^2}$. 
This exponent suppresses the contribution of the points near $s=m_\tau^2$ with low 
experimental accuracy and unknown tail beyond them. However this suppression  is not 
always enough, especially when one would like to find the excess due to operators $O_{6,8}$
over dominating $f_\pi^2$. 

Gaussian sum rules have an advantage, that the high energy tail in the dispersion integrals
are suppressed by the factor $e^{-s^2/M^4}$, stronger than in Borel sum rules even at $M^2$
not much lower $m_\tau^2$. But they also have a disadvantage, because the factorial suppression
of high order terms in OPE starts in fact at operators of very high dimension.

\begin{figure}[tb]
\hspace{5mm}
\epsfig{file=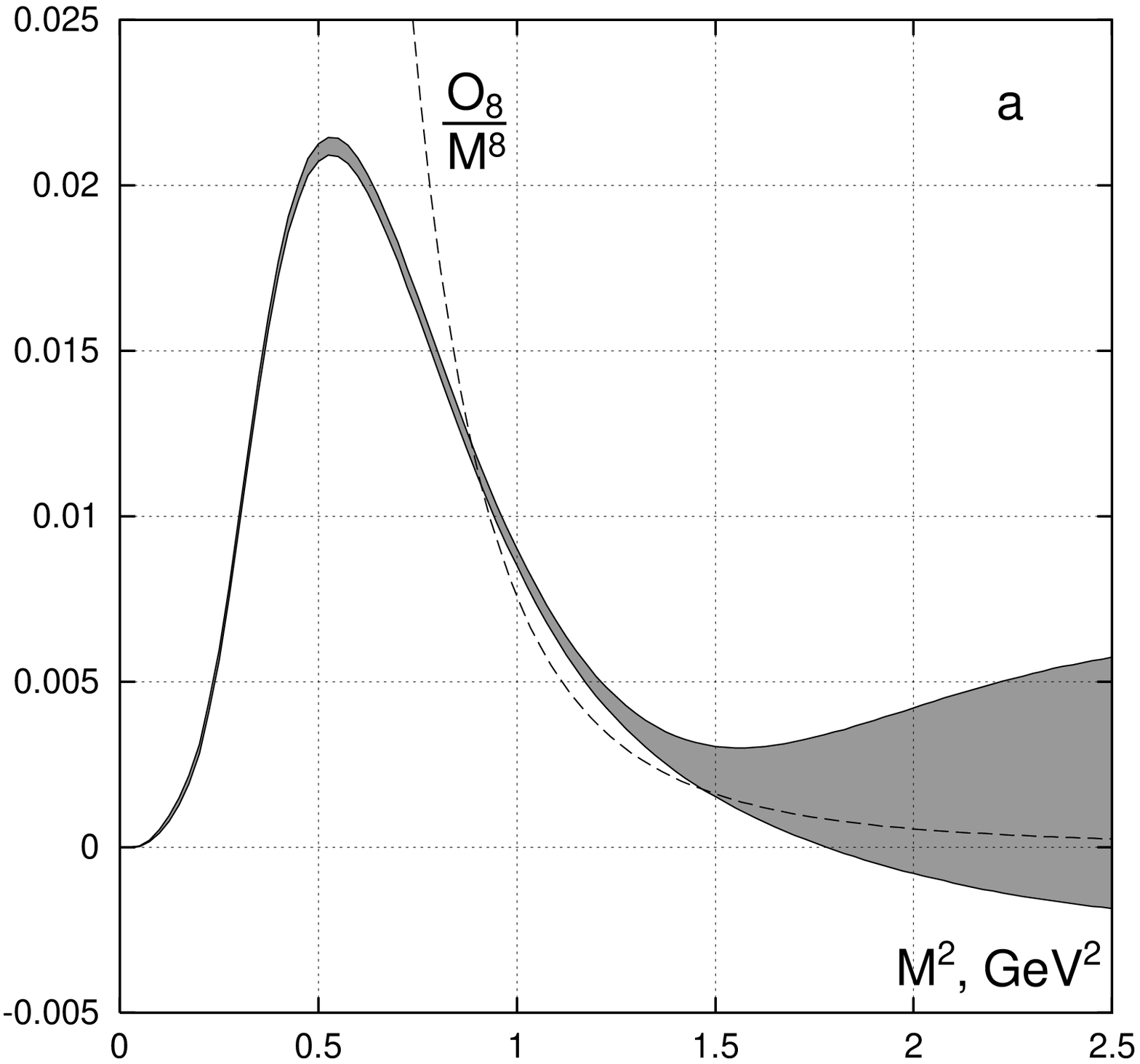, height=65mm, width=70mm} \hspace{10mm}
\epsfig{file=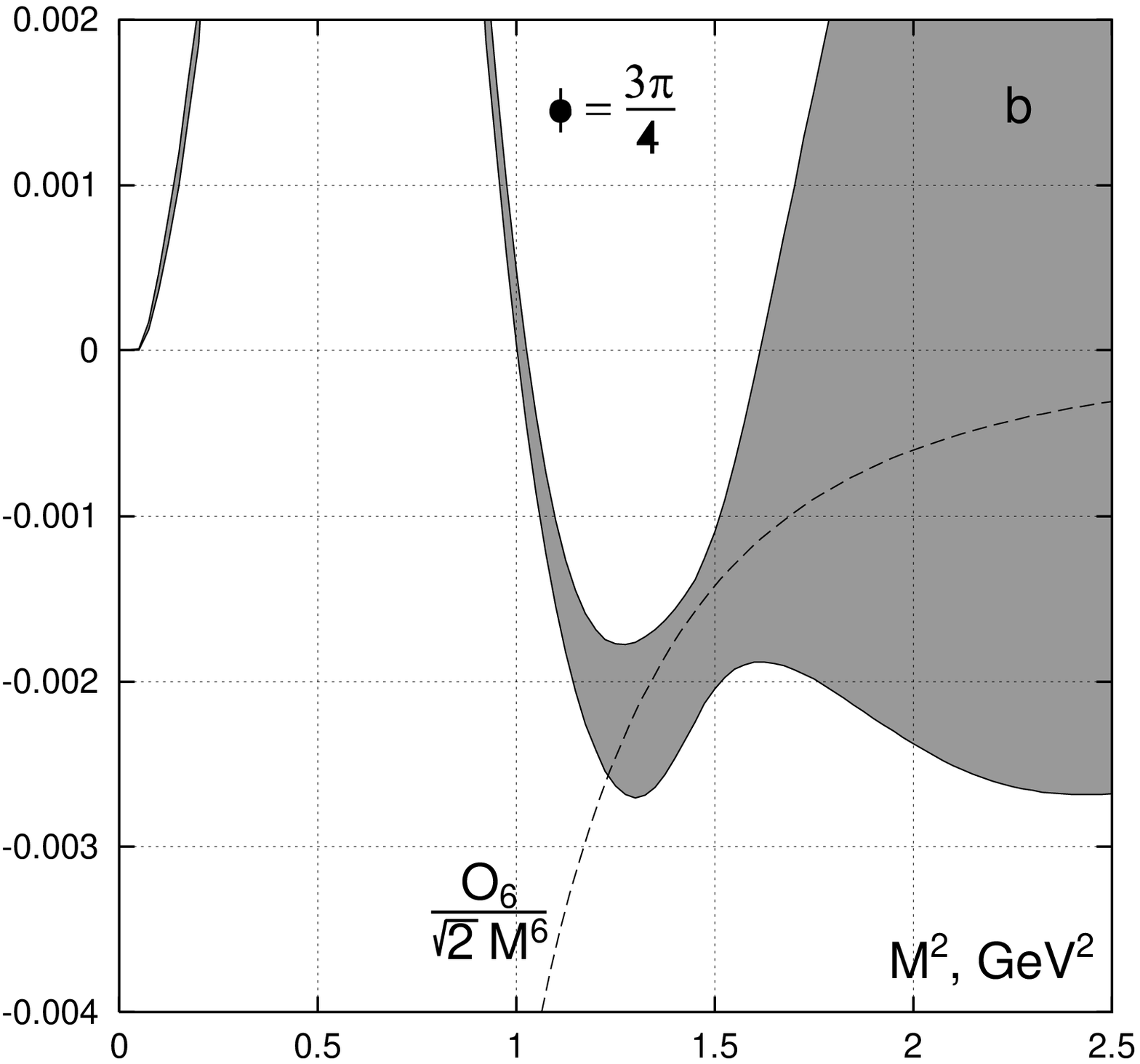, height=65mm, width=70mm}
\caption{Left hand sides of the equations (\ref{4ksr}) and (\ref{4kp2sr}) respectively.
Dash lines display OPE prediction with operators taken from Borel sum rules (\ref{borbest}).}
\label{gausso6o8}
\end{figure}

The sum rules of this kind can be constructed with the help of the analysis of the correlators
on the complex plane. Consider for instance the real part of the polarization operator
on the imaginary axes:
\be
{\rm Re}\,\Pi^{(1)}_{V-A}(is_0)\,=\,{1\over 2\pi^2}\,\int_0^\infty 
{(v_1-a_1)(s)\over s^2+s_0^2}\,s\,ds\,=\,\sum_{k=1}^\infty {O_{4k}\over (-s_0^2)^k}
\ee
Since both sides of this equation depend only on $s_0^2$, one may apply the Borel operator
over this variable to get the following gaussian sum rule:
\be
\label{4ksr}
{1\over 2\pi^2}\,\int_0^\infty 
e^{-s^2/M^4}(v_1-a_1)(s)\,{s\,ds\over M^4}\,=\,\sum_{k=1}^\infty 
{(-)^k O_{4k}\over (k-1)!\, M^{4k}}
\ee
Since the operator $O_4$ is negligible, the expansion in the r.h.s. starts from $O_8$. 
Consequently the eq.~(\ref{4ksr}) can be used to find the operator of dimension 8. 
The l.h.s. of (\ref{4ksr}) is plotted in Fig.~\ref{gausso6o8}a.

In order to find the operator $O_6$ from Gaussian-like sum rule, one has to construct an another
function of $s_0^2$ from the correlator $\Pi(s)$ which consists of the operators of
dimension $4k+2$. To kill the leading term $f_\pi^2$, consider the imaginary part
of $e^{i\varphi}\Pi(s_0e^{i\varphi})$ at some angle $\varphi$. 
Further, to cancel the operators of dimension
$4k$, one can add this function at the point symmetric with respect to imaginary axes. The result is:
$$
{1\over 2s_0}\,{\rm Im}\left[ \,e^{i\phi/2}\Pi^{(1)}_{V-A}(s_0e^{i\phi/2})
+e^{-i\phi/2}\Pi^{(1)}_{V-A}(-s_0e^{-i\phi/2}) \,\right]\,=
$$
\be
=\,{\sin{\phi}\over 2\pi^2}\int_0^\infty \, {(v_1-a_1)(s) \over s^4+s_0^4 - 2s^2s_0^2\cos{\phi}}\,s^2\,ds
\,=\,\sum_{k=1}^\infty \,\sin{(k\phi)}\,{O_{4k+2}\over s_0^{2k+1}}
\ee
Applying the Borel operator by variable $s_0^2$, we get:
\be
\label{4kp2sr}
{1\over 2\pi^2}\int_0^\infty 
\exp{\!\left({s^2\over M^4}\cos{\phi}\right)}\,\sin{\!\left({s^2\over M^4}\sin{\phi}\right)}
(v_1-a_1)(s)\,{ds\over M^2}\,=\,\sum_{k=1}^\infty 
\sin{(k\phi)} \,{O_{4k+2}\over k!\, M^{4k+2}}
\ee
The r.h.s. starts from $O_6$. For the exponent to be decreasing, the angle $\phi$ must be in the
range $\pi/2<\phi<\pi$ (the range $\pi<\phi<3\pi/2$ obviously does not contain any new information). 
The l.h.s. of (\ref{4kp2sr}) for $\phi=3\pi/4$ is plotted
versus $M^2$ on Fig.~\ref{gausso6o8}b.

The operators $O_{4k}, k\ge 1$ contribute to the sum rule (\ref{4ksr}). If we neglect contributions 
of high orders starting from $O_{12}$, then from the Fig \ref{gausso6o8}a at $M^2=1.5 \, {\rm GeV}^2$
for $O_8$ we would have
\be
O_8\,=\,(10.6\pm 3.6)\times 10^{-3} \; {\rm GeV}^8
\label{o8gn1}
\ee
However, the result strongly depends on $O_{12}$ operator. If we use the same estimation as
in the previous section $|O_{12}|=5|O_6|$, then (\ref{o8gn1}) may change on $\pm 7.7\times 10^{-3}$.
Considering this amount as possible error in (\ref{o8gn1}) and adding the errors in 
quadratures, we get:
\be
O_8\,=\,(10.6\pm 8.5)\times 10^{-3} \; {\rm GeV}^8
\label{o8gn2}
\ee
Going to lower energies is dangerous, because the contribution of $O_{12}$ increases drastically.
At higher $M^2$, where the higher order operator can be neglected, the experimental error
does not allow to get any definite conclusion about the magnitude of $O_8$.

Now we turn to eq.~(\ref{4kp2sr}) at $\phi=3\pi/4$. The most suitable scale is 
$M^2=1.5 \; {\rm GeV}^2$. The next to $O_6$ in (\ref{4kp2sr}) is the operator $O_{10}$. 
Its contribution at $1.5\, {\rm GeV}^2$ is suppressed not so much, by a factor $(\sqrt{2}M^4)^{-1}$.
If we allow, that $|O_{10}|$ could be as large as $2|O_6|$ (in GeV) and include this uncertainty
as an error, then the following esimation goes from Fig \ref{gausso6o8}b:
\be
O_6\,=\,-\,(7.2\pm 5.1)\times 10^{-3}\; {\rm GeV}^6
\label{o6gn1}
\ee

In case of more optimistic assumption used in previous section $|O_{10,12}|\sim |O_6|$
we get better results, especially for the operator $O_8$:
\be
\label{gaussbest}
O_6\,=\,-(7.7\pm 3.2)\times 10^{-3} \; {\rm GeV}^6 \; , \qquad
O_8\,=\,(9.8\pm 2.3)\times 10^{-3} \; {\rm GeV}^8
\ee
One may stress, however, that the assumption $|O_{12}|\sim |O_6|$ is the most dubious.

The conclusion is the following. The operators, obtained from Gaussian sum rules are compatible
with Borel ones. However the range of applicability is different. Indeed, the Borel exponent
effectively suppresses the high energy contribution only for $M^2<1\, {\rm GeV}^2$. 
But in the Borel expansion each operator of dimension $D$ has the factor $1/(D/2)!$, which 
provides much stronger suppression then the Gaussian factor $1/(D/4)!$. 
Consequently the effective radius of convergence of borel series could be lower. 
As the Figs \ref{borc}-\ref{bors}
show, this is indeed true: the coincidence of the right and left hand sides begins with
$M^2=0.6\,{\rm GeV}^2$, twice lower the correspondent gaussian value.

\section{Summary}

The recently obtained data by ALEPH and OPAL collaborations on $V-A$ spectral functions
in $\tau$-decay were used for determination of quark and quark-gluon condensates:
vacuum expectation values of dimension 6 and 8 operators $O_6^{V-A}$ and $O_8^{V-A}$.
The analytical properties of polarization operator $\Pi^{(1)}_V(q^2)-\Pi^{(1)}_A(q^2)$
in the complex $q^2$-plane were exploited. Three types of sum rules were used:
moments sum rules, Borel and Gaussian ones. The results are summarized in Tables 1,2.
They are in agreement with one another in the limit of errors and the best values of $O_{6,8}$ are:
\be
O_6\,=\,-(6.8\pm 2.1)\times 10^{-3} \; {\rm GeV}^6 \; , \qquad
O_8\,=\,(7\pm 4)\times 10^{-3} \; {\rm GeV}^8
\label{sumval}
\ee
The errors here are not quite well defined, they are just our estimations based on the data, 
presented in Tables 1,2. Particularly, in case of $O_8$ operator the errors strongly depend
on the assumption about the magnitude of $O_{12}$. In the most pessimistic case of large $|O_{12}|$,
say $|O_{12}|\sim 5|O_6|$ in GeV, we have only the upper limit 
$O_8\lesssim 14\times 10^{-3}\, {\rm GeV}^8$.

The values (\ref{sumval}) are by a factor $1.5-2$ larger, than the values (\ref{op6e1}), (\ref{op8e1})
found from low energy theorems and QCD sum rules (see Introduction). If this discrepancy
is addressed to the quark condensate, then, in accord with (\ref{qc1}) it 
means that $m_u+m_d$ at $1 \, {\rm GeV}^2$
is by $20-40\%$ less than the standard value $12\, {\rm MeV}$. Up to this may be not so essential
discrepancy the analysis of $\tau$-decay data confirms the general concept of OPE and the
magnitudes of quark and quark-gluon condensates.

\begin{table}
\hspace{10mm}
\begin{tabular}{c|c|c||c||c|c|c|c}
\hline
{\small source} & {\small assumption} & {\small scale $M^2$} & {\small central} & 
 {\small exp.} & {\small $3\%$ $f_\pi^2$} & {\small $O_{10}$} & {\small total} \\ 
           &  &{\small in GeV${}^2$}  & {\small value of $O_6$} & 
{\small error} & {\small error} & {\small error} &  {\small error} \\  \hline\hline 
{\small eq.~(\ref{bre}) at}  & {\small $O_{10}\sim 2O_6$} &
0.80 & $-6.3$ & 1.4 & 1.3 & 1.6 & 2.5 \\  \cline{2-8}
{\small $\phi=5\pi/6$} & {\small $O_{10}\sim O_6$} & 
0.62 & $-6.1$ & 0.7 & 0.8 & 1.3 & 1.7 \\  \hline
{\small eq.~(\ref{bim}) at}  & {\small $O_{10}\sim 2O_6$} & 
0.85 & $-6.8$ & 1.4 & --- & 1.6 & 2.1 \\  \cline{2-8}
{\small $\phi=2\pi/3$} & {\small $O_{10}\sim O_6$} & 
0.85 & $-6.8 $& 1.4 & --- & 0.8 & 1.6 \\  \hline
{\small eq.~(\ref{4kp2sr}) at}  & {\small $O_{10}\sim 2O_6$} & 
1.53 & $-7.2$ & 2.8 & --- & 4.3 & 5.1 \\  \cline{2-8}
{\small $\phi=3\pi/4$} & {\small $O_{10}\sim O_6$} & 
1.47 & $-7.7$ & 2.0 & --- & 2.5 & 3.2 \\  \hline
\end{tabular}
\caption{Values of the operator $O_6$ with possible errors in $10^{-3}\, {\rm GeV}^6$, 
obtained from Borel and Gaussian sum rules. In each case the scale $M^2$ is choosen in such way,
that the total squared error (the sum of all squared errors), is minimal. In second column the
magnitudes of operators are given in GeV.}
\end{table}

\begin{table}
\hspace{4mm}
\begin{tabular}{c|c|c||c||c|c|c|c|c}
\hline
{\small source} & {\small assumption} & {\small scale $M^2$} & {\small central} & 
 {\small exp.} & {\small $3\%$ $f_\pi^2$} & {\small $O_6$} & {\small $O_{10, 12}$} & {\small total} \\ 
           &  &{\small in GeV${}^2$}  & {\small value of $O_8$} & 
{\small error} & {\small error} & {\small error} & {\small error} &  {\small error} \\  \hline\hline 
{\small eq.~(\ref{bre}) at}  & {\small $O_{10}\sim 2O_6$} &
0.65 & 6.2 & 2.8  & 1.2 & --- & 7.4 & 8.0 \\  \cline{2-9}
{\small $\phi=3\pi/4$} & {\small $O_{10}\sim O_6$} & 
0.59 & 5.5 & 1.3 & 0.9 & --- & 4.1 & 4.4 \\  \hline
{\small eq.~(\ref{bim}) at}  & {\small $O_{12}\sim 5O_6$} & 
0.65 & 7.0 & 1.0  & --- &  5.8  & 4.0 & 7.1 \\  \cline{2-9}
{\small $\phi=3\pi/4$} & {\small $O_{12}\sim O_6$} & 
0.41 & 7.2 & 0.1 & --- & 2.8 & 2.0 & 3.4 \\  \hline
{\small eq.~(\ref{4ksr})}  & {\small $O_{12}\sim 5O_6$} & 
1.49 & 10.6 & 3.6  & --- & ---  & 7.7 & 8.5 \\  \cline{2-9}
   & {\small $O_{12}\sim O_6$} & 
1.29  & 9.8 & 1.0 & --- & --- & 2.0 & 2.3 \\  \hline
\end{tabular}
\caption{Values of the operator $O_8$ with possible errors in $10^{-3}\, {\rm GeV}^8$.}
\end{table}

\vspace{10mm}

The authors are thankful to A.Oganesian for his help in the calculation of dimension 8 operator.
The research described in this publication was made possible in part by Award No RP2-2247
of U.S. Civilian Research and Development Foundation for Independent States of Former
Soviet Union (CRDF) and by Russian Found of Basic Research grant 00-02-17808.

\section*{Appendix: The condensate of dimension 8\footnote{
A.Oganesian participated in the calculation of dimension 8 operator.}}

It consists of three different contributions:
\be
O_8\,=\,O_8^{(0)}\,+\,O_8^{(2)}\,+\,O_8^{(4)} \; ,
\ee
where the upper index denotes the number of quarks in vacuum. The purely gluonic condensate
$O_8^{(0)}$ and two-quark one $O_8^{(2)}$ have been computed in \cite{BG}. They contain
many independent operators, which cannot be expressed in terms of condensates of lower dimensions.
However in the masseless quark limit these operators are the same for vector and axial
correlators and disappear in the difference $O_8^{V-A}$.

We have explicitly computed the four quark condensate for the vector
current correlator:
\bea
O^{V\,(4)}_8 &  = & 
{g^2\over 36 }\, \Bigl<
\,8\,(\bar{u}\gamma_\alpha \lambda^a\! \stackrel{\leftrightarrow}{D}_\beta d)
(\bar{d} \gamma_\alpha  \lambda^a\! \stackrel{\leftrightarrow}{D}_\beta u) \,
-\,11\,g\,f^{\,abc}G_{\alpha\beta}^c
(\bar{u} \gamma_\alpha\gamma_5 \lambda^a d)(\bar{d} \gamma_\beta\gamma_5  \lambda^b u)
\nonumber \\ 
 & &
-\,14\, (\bar{u}\overleftarrow{D}_\alpha \gamma_\beta\gamma_5 
\lambda^a \overrightarrow{D}_\alpha d )
(\bar{d}\gamma_\beta\gamma_5 \lambda^a u)
\,-\,14\,(\bar{d}\,\overleftarrow{D}_\alpha \gamma_\beta\gamma_5
\lambda^a \overrightarrow{D}_\alpha u)
(\bar{u}\gamma_\beta\gamma_5 \lambda^a d)
\nonumber \\ 
 & &
+\,(\bar{u}\gamma_\alpha \{ \tilde{G}_{\alpha\beta}, \lambda^a\}d)
(\bar{d}\gamma_\beta\gamma_5 \lambda^a u) 
\,+\,(\bar{d}\gamma_\alpha\{ \tilde{G}_{\alpha\beta}, \lambda^a\}u)
(\bar{u}\gamma_\beta \gamma_5 \lambda^a d)\, \Bigr> \; ,
\label{4qvecall} 
\eea
where  $G_{\alpha\beta}={g\over 2}\lambda^a G^a_{\alpha\beta}$ is gluon field,
${\tilde G}_{\alpha\beta}={1\over 2}\ve_{\alpha\beta\mu\nu}G_{\mu\nu}$, 
$\ve_{0123}=1$, $\gamma_5=i\gamma_0\gamma_1\gamma_2\gamma_3$,
$\stackrel{\leftrightarrow}{D}=\overrightarrow{D}-\overleftarrow{D}$,
$D_\mu=\partial_\mu-{ig\over 2}\lambda^a A^a_\mu$,
the derivatives in the brackets $(\ldots)$ act only on the objects inside these brackets;
$\lambda^a$ are Gell-Mann matrices, $[\lambda^a,\lambda^b]=if^{abc}\lambda^c$.
 One may check, that the eq.~(\ref{4qvecall}) can be brought to the form obtained 
in \cite{GP}, which verifies our results.

The condensate of axial currents $O^{A\,(4)}_8$ can be easily obtained from 
(\ref{4qvecall}) with the help of the replacement $d\to \gamma^5 d$. 

To reduce the number of independent
operators in (\ref{4qvecall}), the vacuum insertion can be applied to $O^{(4)}_8$. 
Nevertheless this procedure is not unambiguous. Indeed, let us consider the following
operator, which appears in the derivation of (\ref{4qvecall}):
\bea
<(\bar{q}\gamma_\alpha\lambda^a q) D_{(\alpha}D_{\beta)}(\bar{q}\gamma_\beta\lambda^a q)>
 & = & {g\over 4}<f^{abc}G^c_{\alpha\beta}(\bar{q}\gamma_\alpha\lambda^a q)
(\bar{q}\gamma_\beta\lambda^b q)>  \nonumber \\ 
 & = & -\,{i\over 2}<\bar{q}\hat{G} q><\bar{q}q> \; ,
\label{viex1}
\eea
where $\hat{G}=\gamma_{\alpha\beta}G_{\alpha\beta}$. In (\ref{viex1}) we used the
quark equation of motion and commutational relation for the covariant derivatives, and then
applied   the vacuum insertion. On the other hand, one may apply the vacuum insertion at first
and use equations of motion after then:
\bea
<(\bar{q}\gamma_\alpha\lambda^a q) D_{(\alpha}D_{\beta)}(\bar{q}\gamma_\beta\lambda^a q)>
 & = & -\left( 1-{1\over N_c^2} \right) <\bar{q} D^2 q><\bar{q}q> \nonumber \\ 
 & = &  -\, {i\over 2}\left( 1-{1\over N_c^2} \right)
<\bar{q}\hat{G} q><\bar{q}q> \; ,
\label{viex2}
\eea
where $N_c$ is the number of colors. We see, that two different ways of the vacuum insertion
give the same result only up to the terms of order $\sim 1/N_c^2$.

Consequently, within the framework of the factorization hypothesis we 
may write the four quark operators (\ref{4qvecall}) in  the following  form:
\be
O^{V\,(4)}_8\,=\,-\,O^{A\,(4)}_8\,=
\,4\pi\alpha_s m_0^2 <\bar{q}q>^2 \left( \,1\,+\,O(N_c^{-2})\, \right)
\label{4qvacf}
\ee
The parameter $m_0$ is introduced as $<\bar{q}\hat{G}q>=im_0^2<\bar{q}q>$ according to
\cite{DS}. Within this accuracy the equation (\ref{4qvacf}) coincides with the result of
ref.~\cite{DS} (according to the isotopic symmetry the current correlators
$\Pi_{\mu\nu}^{\rm (here)}=2 \Pi_{\mu\nu}^{\mbox{\rm\scriptsize (ref.~\cite{DS})}}$).

\end{document}